\documentclass[aps,epsfig,twocolumn]{revtex4}

\usepackage{graphicx}
\usepackage{dcolumn}
\usepackage{bm}

\newcommand{\bq}{\begin{equation}}
\newcommand{\eq}{\end{equation}}
\newcommand{\bqa}{\begin{eqnarray}}
\newcommand{\eqa}{\end{eqnarray}}
\newcommand{\nn}{\nonumber \\}

\begin{document}
\draft
\title{ 
Observation of the emergent photon in a metastable fractionalized phase of 
Bose-condensed excitons: Monte Carlo simulation
}

\author{Sung-Sik Lee and Patrick A. Lee}
\address{Department of Physics, Massachusetts Institute of Technology,\\
Cambridge, Massachusetts 02139, U.S.A.\\
}
\date{\today}
       
\begin{abstract}

Using Monte Carlo simulation, 
we studied fractionalized phases
in a model of exciton Bose condensate
and found evidence for an emergent photon 
in a finite size system.
The fractionalized phase is a meta-stable Coulomb phase
where an emergent photon arises as a gapless collective
excitation of the excitons.
We also studied a possibility of 
spiral$^*$ phase where fractionalization 
and long range spiral order of the exciton condensate coexist.

\end{abstract}
\maketitle

\section{Introduction}
Recently, possible exotic phases in exciton Bose condensate 
have been studied in the strong coupling regime\cite{LEE1}.
While the Bose condensation of exciton is an interesting subject by itself\cite{BUTOV},
it turns out that this system can serve as a good theoretical laboratory to study fractionalization.
Fractionalized phase is a many-body state where
the emergent low energy excitation carries 
fractional quantum number of the microscopic degree of freedom\cite{EXCEPTION}.
Since the Anderson's proposal of the resonating valence bond state\cite{ANDERSON},
several concrete models have been proposed to show fractionalization, 
such as
exact solvable spin models\cite{KITAEV,WEN2003PRL},
dimer model\cite{MOESSNER},
bosonic models\cite{MOTRUNICH,WEN2002PRL,MOTRUNICH2004,LEE1},
and frustrated spin model\cite{HERMELE}.
Fractionalization can be naturally described in terms of 
deconfinement phase of gauge theory.
The gauge group can be discrete
like the $Z_2$ gauge theory\cite{KITAEV,WEN2003PRL,MOESSNER}
or it can be continuous like
the U(1) gauge theory\cite{MOTRUNICH,WEN2002PRL,MOTRUNICH2004,HERMELE,LEE1}.
While there exist exactly solvable microscopic models\cite{KITAEV,WEN2003PRL}
which show the fractionalized phase with the $Z_2$ gauge field at low energies,
there exists no such exactly solvable model or numerical verification 
for the existence of the fractionalized phase with the U(1) gauge field.
Therefore it is of great interest to numerically show the occurrence of 
fractionalization in the model of exciton Bose condensate which supports
the U(1) gauge field in the fractionalized phase\cite{LEE1}.
Here we emphasize that currently there exists no
experimental realization of the exciton model\cite{LEE1}.
We take the model as a purely theoretical model to show
the existence of the fractionalized phase as a proof of principle.

We consider the Bose condensation of excitons which are
made of an electron in the a-th conduction band and a hole in the b-th valence band
where $a =1,2,...,N_c$ and $b =1,2,...,N_v$ denote 
the band (flavor) indices. 
In general, the degeneracy of the conduction band $N_c$ and the valence band $N_v$ is different.
Here we will consider the case where $N_c = N_v = N$.
We denote by  $\theta^{ab}$ the phase of the $ab$-exciton field.
When the diagonal ($a = b$) excitons are condensed,
the dynamics of the off-diagonal ($a \neq b$) exciton becomes relativistic
as a result of the dynamical constraint posed by the diagonal excitons.
The off-diagonal exciton is expected to 
have the usual disordered or Bose condensed phases.
Interestingly, we found that
in the strong coupling regime a new phase can arise where
excitons scatter with each other to 
rapidly exchange their constituent particle or hole\cite{LEE1}. 
As a result, the exciton loses its identity at a long distance scale
and a half of exciton which carries only one flavor index 
arises as an effective degree of freedom.
The emergent fractionalized degree of freedom can be either boson or fermion
depending on the coupling constants\cite{LEE1}.
Although the flavor quantum number of the fractionalized particle is 
inherited from the original electron/hole, 
the fractionalized particles are different objects from the original
electron/hole in the sense that they don't carry electric charge.
Instead they are coupled to a new emergent gauge field.
The U(1) gauge theory in (3+1)D can be in deconfinement phase 
although the fractionalized particles are gapped.
The gauge field in the deconfinement phase behaves in the same way as the photon in our world
and has been referred to as emergent photon in literatures. 
The gauge field is, on the other hand, nothing but a low energy 
collective excitations of the exciton.
In the space-time picture, the world sheet of the electric flux line 
for the emergent gauge field corresponds to the web formed by
strongly interacting exciton world lines (world line web)\cite{LEE1}.
The boundary of the world line web becomes the world line of the fractionalized particle. 
This explains why the gauge field should emerge as a result of fractionalization\cite{LEVIN}.

It has been pointed out that fractionalization can also occur 
in the presence of conventional long range order 
(dubbed as the AF$^*$ phase in the case 
where the long range order is the antiferromagnetic order)
because the condensation of gauge neutral order parameter field 
can leave the deconfinement phase of the emergent gauge theory intact\cite{SENTHIL1}.
This possibility has also been studied in the exciton system\cite{LEE2}.
Although the previous studies\cite{LEE1,LEE2} provide the rather complete picture,
the occurrence of fractionalization in exciton system can be inferred only 
in the large $N$ limit.
It is of great interest to verify the occurrence of fractionalization 
by numerical simulation at some finite $N$. 
This is the purpose of the present paper. 

How do we know if fractionalization occurs ?
The most direct way would be to probe fractionalized excitation.
However, it is difficult to directly observe fractionalized particles
because they are either gapped or
strongly interacting with the gauge field. 
An alternative way is to probe the emergent non-confining gauge field which 
inevitably arises as a consequence of fractionalization.
The effect of emergent gauge field is most dramatic 
because it is gapless in the fractionalized phase. 
The exciton system supports massless U(1) gauge boson in the fractionalized phases\cite{LEE1,LEE2}. 
Currently, there exist other models\cite{WEN2002PRL,MOTRUNICH,MOESSNER,HERMELE} 
which were proposed to support the emergent photon in fractionalized phases. 
However, a numerical observation of the emergent photon has not yet been made in those systems.
In the present study, we report the observation of the emergent photon
in a finite size exciton system from Monte Carlo simulation.

\section{Model and probe for emergent photon}
We consider the exciton system in 3+1D whose Euclidean action is given by\cite{LEE1},
\bqa
S  & = &  \sum_i \Bigl[
\sum_{a<b} \Bigl\{
   -  2 \kappa_0 \cos \left( \theta^{ab}_{i+\Delta \tau} - \theta^{ab}_i \right)    \nn
 && - 2  \kappa_x \sum_{\mu=\hat x, \hat y, \hat z} \cos \left( \theta^{ab}_{i+\mu} - \theta^{ab}_i \right)  \nn
&& -  2 \kappa_x^{'} \sum_{\mu= 2 \hat x,  2 \hat y,  2 \hat z} \cos \left( \theta^{ab}_{i+\mu} - \theta^{ab}_i \right)   
\Bigr\}  \nn
&&  - K_3 \sum_{a, b, c} \cos \left( \theta^{a b}_i + \theta^{b c}_i + \theta^{c a}_i  \right) 
\Bigr].
\label{model}
\eqa
Here $\theta^{ab}_i = - \theta^{ba}_i$ is the phase of the off-diagonal exciton with $a \neq b$
with $i$, site index in 3+1D hypercubic lattice.
The first two terms are the discrete form of the relativistic kinetic energy of the off-diagonal exciton.
$\kappa_0$ and $\kappa_x$ are the phase stiffness in the imaginary time and spatial directions respectively.
In the continuum limit of the imaginary time, 
the ratio $\frac{\kappa_x}{\kappa_0} = \left( \frac{\Delta \tau}{\Delta \tau_0} \right)^2$ 
is determined from the choice of the discrete time step $\Delta \tau$,
where $\Delta \tau_0$ is the time step in which the phase stiffness becomes
identical in space and time.
We can study the quantum system from the Euclidean action with 
discrete time step as far as we choose the time step such that
the temporal correlation length of exciton is
larger than the time step.
The finite number of lattice in the time direction corresponds
to the finite temperature of the quantum system 
and the temperature sets the lowest energy scale we can probe in the simulation.
Taking into account a general time step, we set 
$\kappa_0 =   \left( \frac{\Delta \tau_0}{\Delta \tau} \right) \kappa$ and
$\kappa_x =   \left( \frac{\Delta \tau}{\Delta \tau_0} \right) \kappa$,
where $\kappa$ is the intrinsic coupling constant in the isotropic space-time lattice. 
The third term is a coupling between bosons separated by two units of lattice in space, 
which was absent in the original model\cite{LEE1}. 
The reason for introducing this additional term will be explained shortly.
We also define the intrinsic coupling constant $\kappa^{'}$ for the frustration term as
$\kappa_x^{'} =   \left( \frac{\Delta \tau}{\Delta \tau_0} \right) \kappa^{'}$.
The last term is the potential energy term.
Here we consider the cubic interaction with $K_3 > 0$.
In Ref.\cite{LEE1} we showed that Eq. (\ref{model}) is the effective action 
for the exciton model, 
but it is hard to realize the model in real exciton systems for the following reasons.
First, we are considering the bands with the same degeneracy
for the valence and conduction bands.
Second, our result of this paper is that we need either 
a large number of degeneracy or the introduction of 
frustrated interaction for the fractionalization to occur. 
These conditions are difficult to realize in real material.
Finally, we ignore the dipole-dipole interaction between excitons
which may lead to attraction in three dimension.
However, the more mathematically minded reader can take
Eq. (\ref{model}) as a starting point.
It is simply an extension of the (d+1)D X-Y model to multiple flavors.
For example, the effective action may be realizable in Josephson junction array.

In the strong coupling limit $K_3 = \infty$, the phase of the off-diagonal exciton is constrained to satisfy 
\bq
\theta^{ab}_i = \phi^a_i - \phi^b_i,
\label{dc}
\eq
where $\phi^a_i$ is arbitrary.
We will refer to $e^{i \phi^a_i}$ as slave-boson field.
Then we can use the slave-boson field as our dynamical variables and 
drop the potential energy ($K_3$) term in Eq. (\ref{model}).
Here three remarks are in order. 
First, the replacement of the original exciton field by the slave-boson fields is exact in the strong coupling limit ($K_3 = \infty$)\cite{LEE1}.
The purpose of introducing the slave-boson fields is to parameterize the sub-manifold of finite energy in the full phase space of the original exciton fields.
Second, the decomposition in Eq. (\ref{dc}) does not necessarily imply the occurrence of fractionalization.
Fractionalization may or may not occur depending on other coupling constants.
Third, fractionalization arises only in a long distance scale. 
Microscopically, the slave-boson is always confined within an exciton 
which is the result of infinite bare gauge coupling.
However, the slave-bosons can be regarded as being effectively liberated from exciton 
in a long time (distance) scale as they rapidly exchange their partners 
in the strong coupling limit.
We refer to the long distance effective degree of freedom as fractionalized boson in distinction 
from the microscopic slave-boson. 
In the deconfinement phase, the fractionalized bosons which carry only one flavor(band) quantum number
arise as low energy excitation along with the emergent photon.
We will search for the emergent photon as a signature for the fractionalized phase.
The emergence of the massless gauge boson can be probed from the 
algebraically decaying  correlation function of the gauge field.
While the vector potential $A_\mu$ of the gauge field is not gauge invariant,
the Wilson operator $W_C = e^{i \int_C A_\mu dx^\mu}$ which is defined 
for a closed loop $C$ is gauge invariant and we can express the Wilson operator
in terms of the original exciton fields\cite{LEE1}.
We consider the correlation function between the fluctuations of the Wilson operators
which is defined in the unit plaquette of the x-y plane,
\bq
C_W(r) = Re \frac{1}{V} \sum_i \left< \delta W^{xy \dagger}_{1234}(i+r) \delta W^{xy}_{1234}(i) \right>,
\label{fx}
\eq
where 
\bq
W^{\mu \nu}_{abcd}(i) = e^{i ( \theta^{ab}_i + \theta^{bc}_{i +\mu} + \theta^{cd}_{i+\mu+\nu} + \theta^{da}_{i+\nu} )}
\label{W}
\eq
corresponds to the Wilson operator of the emergent gauge theory
and $ \delta W^{\mu \nu}_{abcd} =  W^{\mu \nu}_{abcd} - <  W^{\mu \nu}_{abcd} >$\cite{LEE1}.
$V$ is the volume of the system.
If the gauge coupling is weak 
the above correlation function measures the flux-flux correlation of the emergent gauge theory.
In our case we expect coupling constant of order unit even in 
the deconfinement phase, and $C_W$ involves higher order flux correlation.
In any case, an algebraically decay will indicate the presence of massless photon
and serves as a signature of fractionalization.

\section{Meta-stable Coulomb phase}
Without $\kappa^{'}$, we found that for $N \leq 12$ a first order phase transition 
occurs from disordered to ordered phases.
These phases are conventional phases where there is no emergent photon.
Fractionalized phase is more likely to occur for a larger $N$.
Here we estimate a critical $N_c$ 
for the occurrence of fractionalized phase 
in the absence of the $\kappa^{'}$ term 
based on the mean-field approach\cite{LEE1}.
The reduced action for the slave-boson are of quartic form of $e^{i \phi^a}$,
which can be decomposed by the Hubbard-Stratonovich transformation\cite{LEE1}.
With the slave-bosons integrated out in the disordered phase, 
the theory reduces to the pure U(1) gauge theory, $ - \beta  \sum_{\mu < \nu} \cos b_{\mu \nu}$
where $b_{\mu \nu}$ is the gauge flux.
The inverse gauge coupling is given by $\beta \sim 2 t^4 N$, where
$t \sim (N-1) \kappa \chi $ is the effective phase stiffness
for the slave-boson with $\chi \sim \sqrt{< e^{i(\theta^{ab}_i - \theta^{ab}_j)} > }$,
the effective hopping integral of the slave-boson.
Here we used $< e^{i(\theta^{ab}_i - \theta^{ab}_j)} > \approx 
< e^{i(\phi^{a}_i - \phi^{a}_j)} > < e^{-i(\phi^{b}_i - \phi^{b}_j)} > = |< e^{i(\phi^{a}_i - \phi^{a}_j)} >|^2 $,
assuming the flavor independent hopping and ignoring the contribution of fluctuations. 
The confinement to deconfinement phase transition occurs 
at $\beta_c \sim 1$ for the 3+1D pure U(1) gauge theory\cite{CRUETZ}.
For $\kappa^{'} = 0$, the order to disorder phase transition occurs 
around $t_c \sim 0.15$ which does not strongly depend on $N$.
This implies that the minimum number of flavors for deconfinement phase to occur 
near the phase boundary is $N_c \sim 1000$ which is too large to be simulated in a large lattice.
In order to reduce $N_c$, we need a larger $t_c$.
For this purpose, we introduced the frustrated coupling with $\kappa^{'} < 0$. 
All of the results presented in this paper is for $N=4$.

With increasing the strength of the frustration, 
the critical $t_c$ for the
disorder/order phase transition increases.
If the frustration is strong enough, a new ordered phase arises 
where excitons are `ferromagnetically' correlated within a $2^3$ box in space
and the boxes are `antiferromagnetically' ordered with each other.
In the box ordered phase, the translational symmetry is broken.
For an intermediate strength of frustration,
there occurs a box liquid phase 
where the long range box order disappears
but there still exists strong box correlation at short distance.
This `box liquid phase' is the fractionalized phase where we observe the emergent photon.
However, it is noted that the region of fractionalized phase is very small in the parameter space.
To be specific we set the frustration $\kappa^{'} N = - 0.39$ with $N=4$.
We choose a $4^3 \times 16$ lattice with periodic boundary condition
where the lattice is longer in the temporal direction than 
the three spatial directions.
This asymmetric geometry is chosen because the phase correlation of exciton is stronger 
in the temporal direction owing to the frustration in the spatial directions.
Further we choose the time step so that the effective phase stiffness is smaller in the temporal direction 
than the spatial directions, that is,  $\kappa_x/\kappa_0 = (1/0.6)^2$.
With this choice, the strength of nearest neighbor exciton correlation becomes comparable
for time and spatial directions.
This effectively enables us to study a longer size in imaginary time direction 
without actually increasing the size of lattice. 
For Monte Carlo simulation we used the local Metropolis algorithm.
It is difficult to implement the more efficient cluster algorithm
because of the need to introduce the further neighbor frustrated interaction.

\begin{figure}
        \includegraphics[height=7cm,width=8cm]{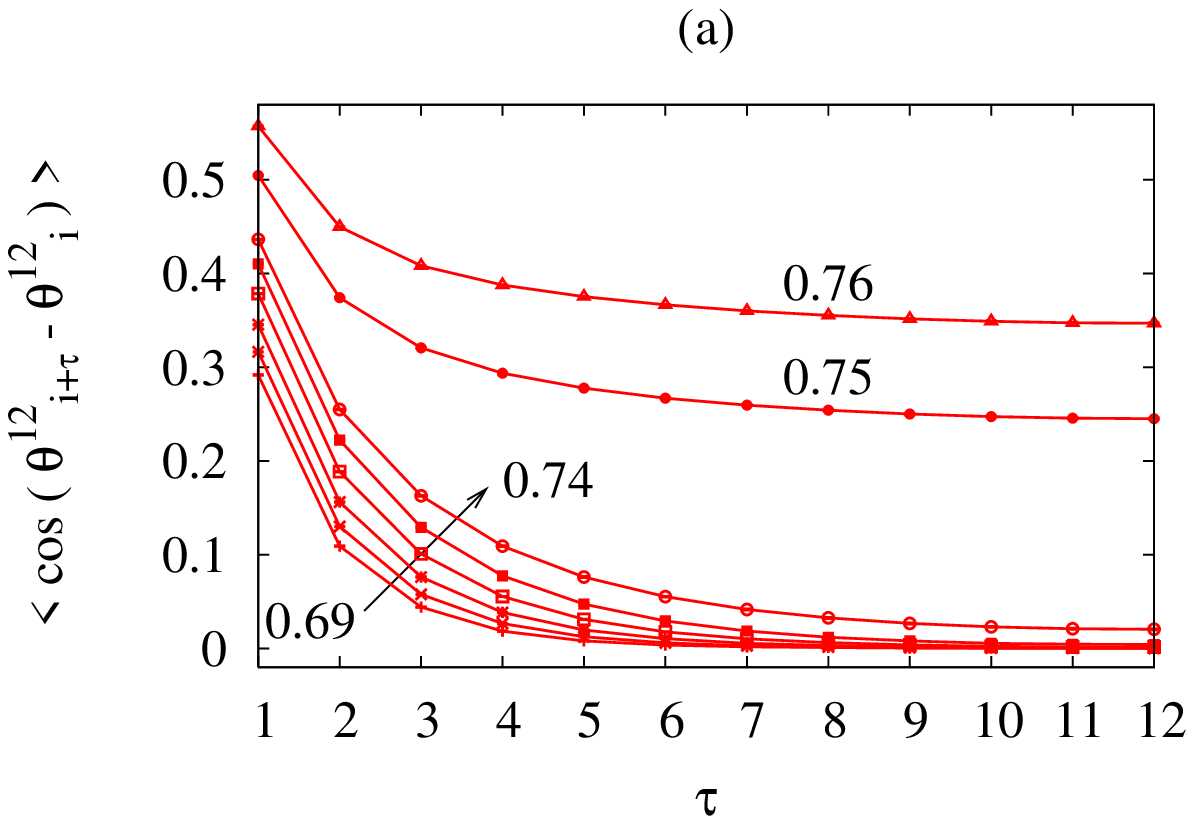}
        \includegraphics[height=7cm,width=8cm]{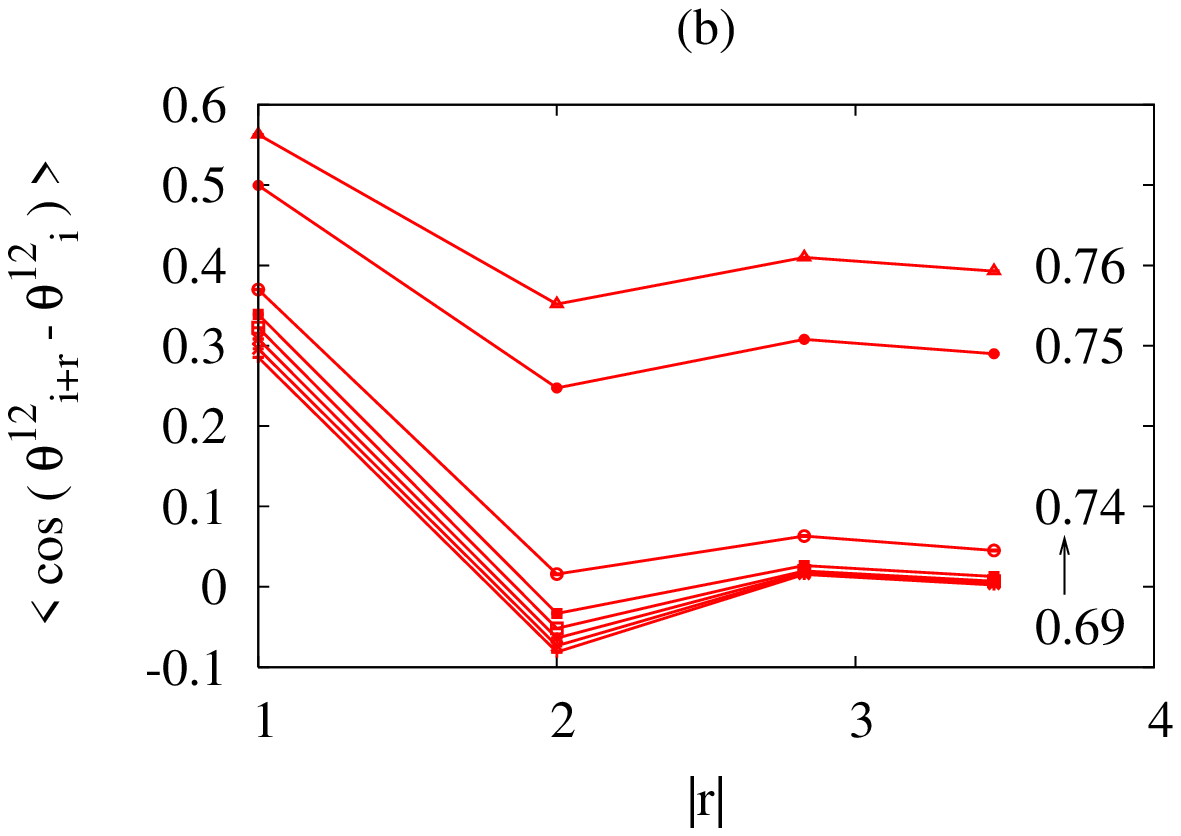}
\caption{
(color online) The correlation function between $\theta^{12}$, the phases of exciton of flavor $1$ and $2$
in the temporal direction (a) and the spatial direction (b).
The average is taken both for the ensemble and for the site $i$.
The numbers in the figure denote the value of $\kappa N$ for each curve.
}
\label{fig:st}
\end{figure}

\begin{figure}
        \includegraphics[height=7cm,width=8cm]{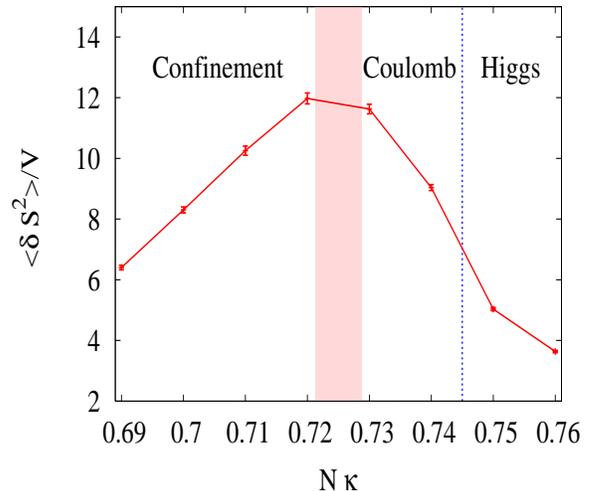}
\caption{
(color online) Mean square of the fluctuation in the action as a function of the phase stiffness
of the off-diagonal exciton.
The dashed line identifies a first order transition between
the disordered phase and the ordered phase.
The shaded region denote the phase transition within the disordered phase
from the confinement phase to the Coulomb phase.
}
\label{fig:c}
\end{figure}

\begin{figure}
        \includegraphics[height=7cm,width=8cm]{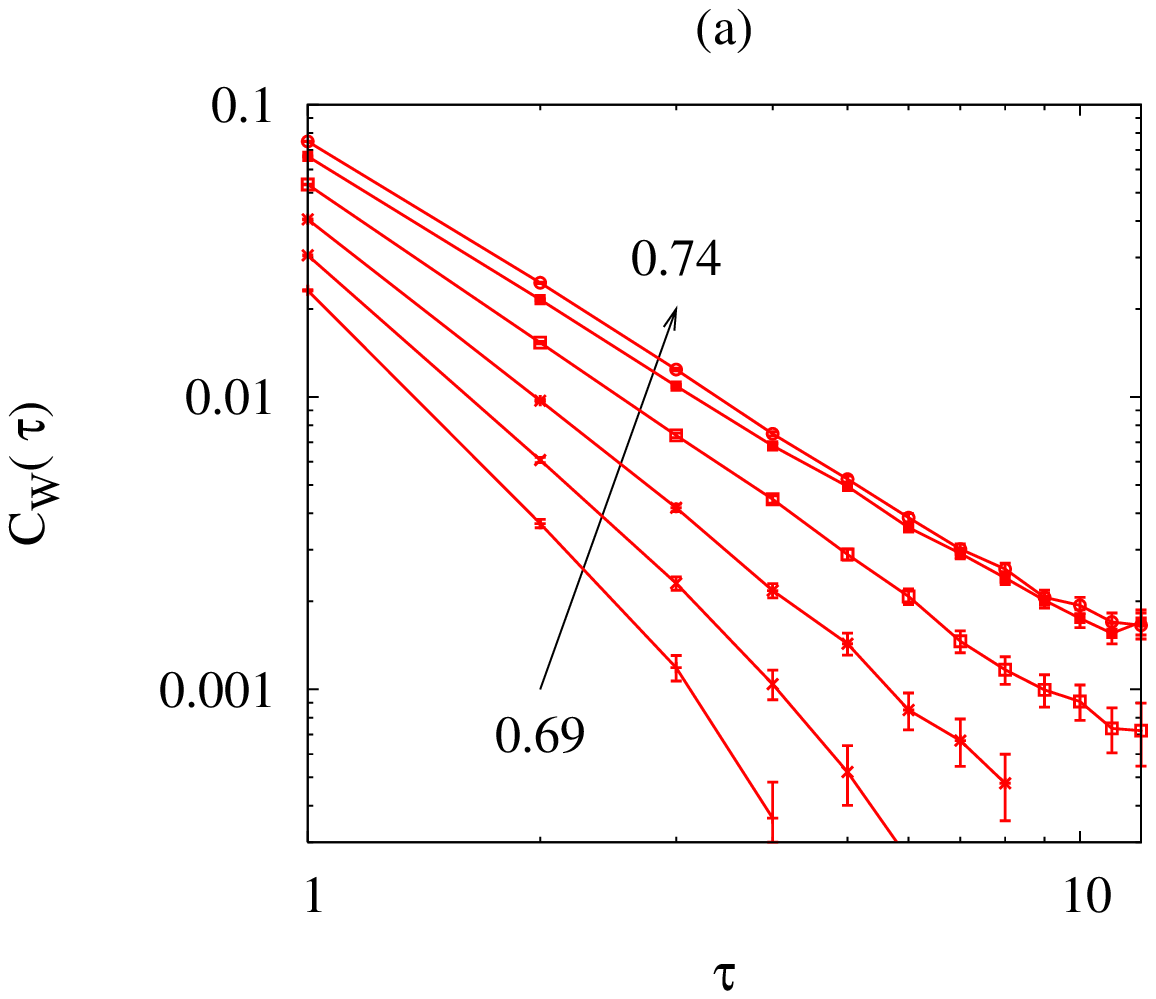}
        \includegraphics[height=7cm,width=8cm]{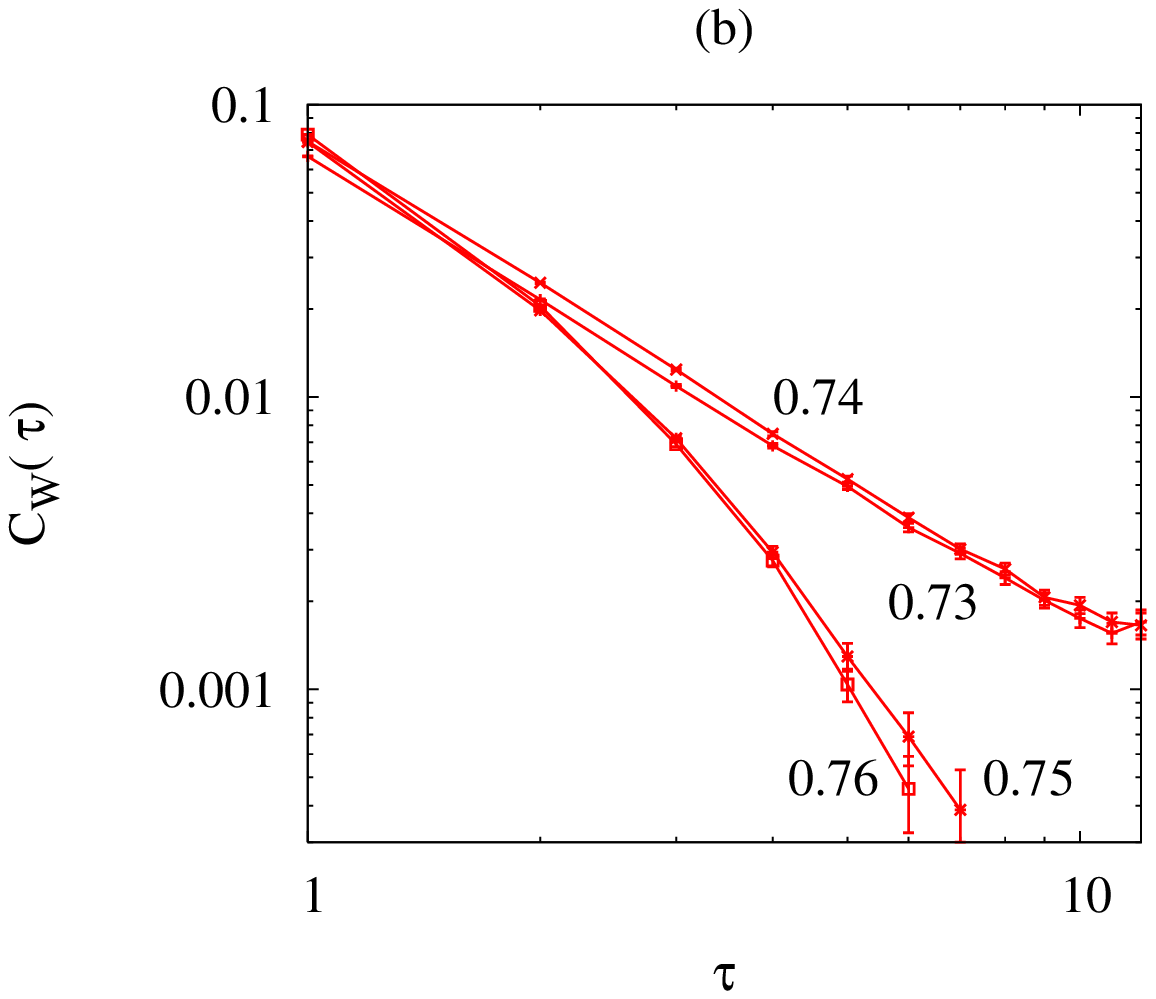}
        \includegraphics[height=7cm,width=8cm]{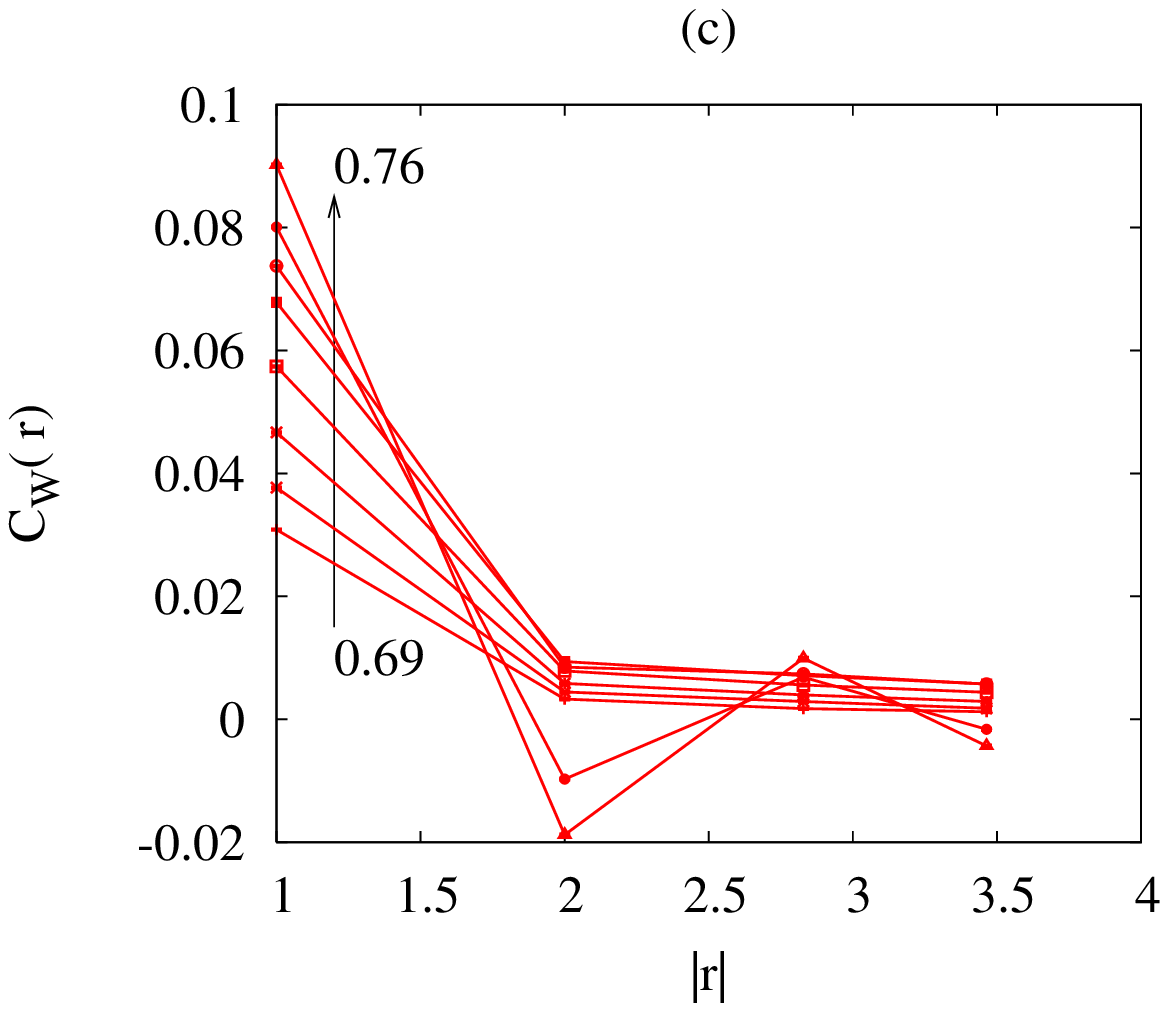}
\caption{
(color online) The correlations function $C_W$ between the fluctuations of Wilson operators.
(a) and (b) are for the temporal direction and (c) is for the spatial direction.
(a) is for $0.69 \leq \kappa N \leq 0.74$ and
(b), for $0.73 \leq \kappa N \leq 0.76$.
}
\label{fig:ft}
\end{figure}

\begin{figure}
        \includegraphics[height=7cm,width=8cm]{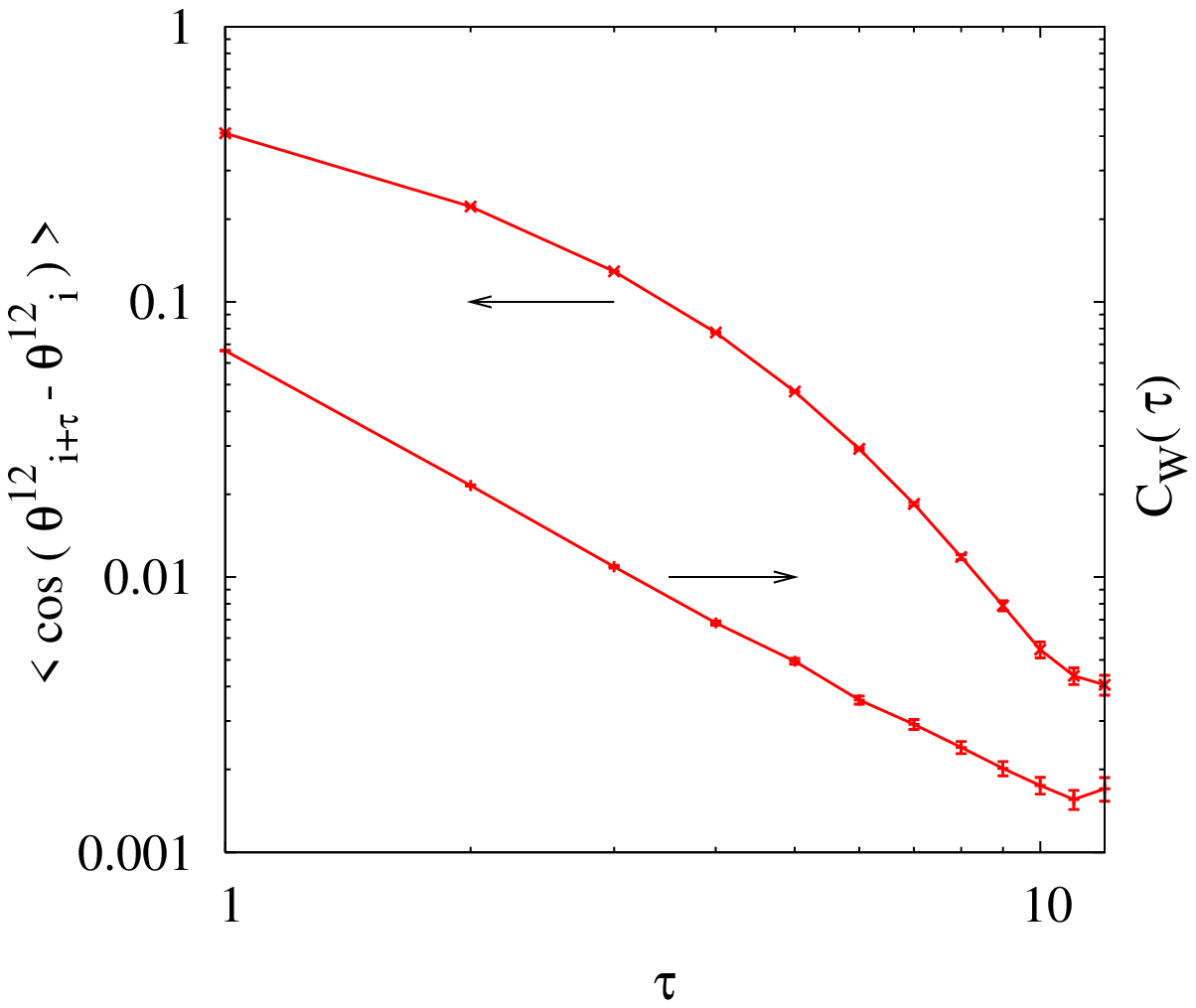}
\caption{
(color online) Comparison between the exciton correlation function and the flux-flux correlation function
in the Coulomb phase.
}
\label{fig:ft_st_0.73}
\end{figure}

In Fig. \ref{fig:st} (a), we display the temporal correlation function of exciton. 
For $\kappa N \leq 0.74$  the correlation function decays exponentially.
This implies that exciton has nonzero mass gap. 
For $\kappa N \geq 0.75$ there exists a long range order and
excitons are Bose condensed. 
The disordered and the Bose condensed phases are separated by 
a first order phase transition. 
The spatial correlation function also exhibits the first order transition 
as is shown in Fig. \ref{fig:st} (b).

In Fig. \ref{fig:c} we show the mean square of fluctuation of the action, $< \delta S^2 >$. 
This quantity is proportional to the ``specific heat'' 
if we interpret the (3+1)D quantum system 
as a 4D classical statistical mechanical system.
A singularity or discontinuity in the ``specific heat'' indicates a quantum phase transition 
in the corresponding quantum system.
The first order phase transition between the disordered phase and the ordered phase
is marked as a discontinuity in the ``specific heat''
between $\kappa N = 0.74$ and $\kappa N = 0.75$.
The interesting thing is that
there is a peak in the ``specific heat'' 
between $\kappa N = 0.72$ and $\kappa N = 0.73$,
suggesting the presence of another phase transition 
within the disordered phase.
In order to understand the nature of two different phases within the disordered phase
we calculate the flux-flux correlation function.
The temporal correlation function $C_W(\tau)$ is shown in Fig. \ref{fig:ft} (a) and (b).
The flux correlation decays faster than power-law for $\kappa N < 0.72$.
Since the absolute value of the correlation function itself is very small, 
it is hard to determine whether it really decays exponentially at long distance.
However it is natural to expect that the correlation function of flux operator
which is composite of gapped exciton fields will also decay exponentially 
unless there is special correlation between excitons.
This phase corresponds to the usual disordered phase of the off-diagonal exciton
where there is no gapless excitations.
In the gauge theory picture, this is the confinement phase
with gapped gauge field. 
What is interesting is the algebraical decay of the flux correlation 
in $0.73 \leq \kappa N \leq 0.74$.
This algebraic decay of $C_W(\tau)$ is purely due to a 
nontrivial correlation among excitons with different flavors
because correlation function of individual exciton is still exponentially decaying.
In order to demonstrate the importance of the correlation,
we compare the exciton correlation function and the flux correlation function
for $\kappa N = 0.73$.
As is shown in Fig. \ref{fig:ft_st_0.73},
the exciton correlation decays exponentially while
the decay of the flux correlation is clearly algebraical.
The power-law decay of $C_W(\tau)$ implies that there exists a gapless mode.
The gapless mode corresponds to the emergent photon.
This phase is the Coulomb phase where the low energy physics
is described by the emergent photon and the gapped fractionalized bosons\cite{LEE1}.
For $\kappa N \geq 0.75$, the flux-flux correlation function decays exponentially again.
This is the usual Bose condensed phase of the off-diagonal excitons.
In this phase, the low energy excitations are the Goldstone modes of the excitons
and there are $(N-1)$ Goldstone bosons associated with the $U(1)^{N-1}$ symmetry
of the model (\ref{model}).
In the gauge theory picture, this phase corresponds to the Higgs phase where
the gauge field becomes gapped due to the Anderson-Higgs mechanism.
One of the $N$ slave-boson modes is `eaten' by the longitudianl gauge field
and $(N-1)$ massless modes are left.
This is consistent with the number of Goldstone modes counted from the original exciton model.

The boundary between the confinement phase and the Coulomb phase is not clearly
distinguished from $C_W(\tau)$.
The flux-flux correlation function decays almost algebraically for $\kappa N = 0.71$ and $0.72$ as well. 
This can be attributed to a cross over behavior at short distance.
If the confining scale is larger than the system size, the flux-flux correlation function
will decay algebraically at short distance even in the confinement phase.
Although not shown here, the absence of a discontinuity 
in the expectation value of the action within our error bar 
suggests that the phase transition from the confinement 
to the Coulomb phases is continuous or of a weak first order.

The observed decaying power of $C_W(\tau)$ in the Coulomb phase is about $2$.
In the pure U(1) gauge theory, the temporal flux-flux correlation function is given by
$ < F_{xy}({\bf r}, \tau) F_{xy}({\bf r}, 0) > 
\sim \sum_{k,\omega} \frac{k_x^2 + k_y^2}{ \omega^2 + (ck)^2 } e^{i \omega \tau}$,
where 
$F_{\mu \nu}$ is the field strength tensor and $c$ is the photon velocity.
In an infinite (3+1)D system, the summations become integral and 
the correlation function goes like $\sim \frac{1}{\tau^4}$.
The discrepancy between the observed decaying power and the predicted power 
may be due to the finite size effect 
associated with the small size in the spatial directions.
With the small size in the space directions,
the discreteness of momentum is important and small momentum will 
dominate the momentum summation.
In an extreme limit, only the smallest possible nonzero momentum such as ${\bf k}=(2\pi/L,0,0)$ 
will contribute to the correlation function.
If the frequency summation is substituted with integral, the correlation function
behaves as $\frac{e^{- m \tau}}{\tau}$ with $m \sim 2 c \pi/L$.
In reality, the decaying power will be between $1$ and $4$ 
with an effective `mass' smaller than $m \sim 2 c \pi/L$.
However, it is hard to make a quantitative prediction 
for the decaying power and the effective `mass'
because this analysis is based on the Maxwellian action in the continuum. 
It is expected that non-Maxwellian higher order terms of the flux field 
and the contribution of large Wilson loops will be also important at short distance.
Despite the small size in the spatial direction, 
we note that the algebraic decay of the $C_W(\tau)$ is not due to the
critical behavior associated with one dimensionality.
If it were due to the one-dimensional effect, 
the exciton correlation function 
should show the power-law decay as well.

In Fig. \ref{fig:ft} (c), the spatial flux-flux correlation function is shown.
The flux correlation decays more slowly in the Coulomb phase compared to the
confinement phase and the Higgs phase.
However, it is hard to see a power-law behavior of the flux correlation function
in the Coulomb phase because of the small lattice size in the spatial direction. 
The oscillatory behavior of the flux correlation in the spatial direction 
is due to the further neighbor frustration  $\kappa^{'}$.

\begin{table}[h]
\begin{tabular}{|c|c|c|}
 \hline
direction & exciton with spiral order & slave-boson with spiral order \\
 \hline
 \hline
x, y & $\theta^{13}$, $\theta^{14}$, $\theta^{23}$, $\theta^{24}$  & $\phi^{1}$, $\phi^{2}$ \\
 \hline
z & $\theta^{12}$, $\theta^{13}$, $\theta^{24}$, $\theta^{34}$  & $\phi^{2}$, $\phi^{3}$\\
 \hline
\end{tabular}    
\caption{
(color online) The second column represents the exciton fields which exhibit spiral long range order in each direction
and the third column, the slave-boson fields which exhibit spiral order in the gauge with $\phi^4=0$.
}
\label{tbl:1}
\end{table}

\begin{figure}
        \includegraphics[height=7cm,width=8cm]{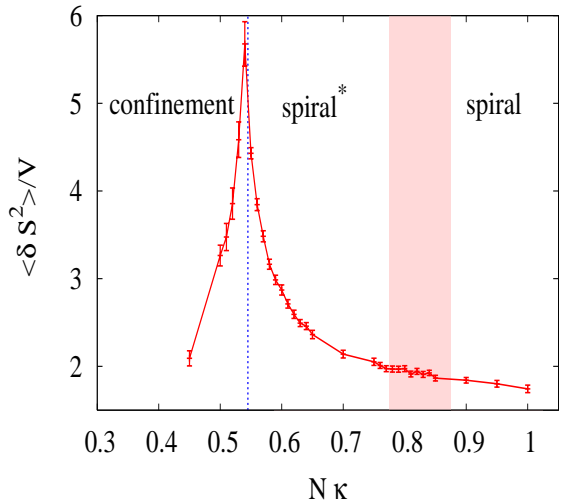}
\caption{
(color online) Mean square of the fluctuation in the action as a function of the phase stiffness
of the off-diagonal exciton.
The dashed line identifies a first order transition between
a disordered (confined) phase and the possible spiral$^*$ phase.
The spiral$^*$ phase is a new phase with spiral order
in $\theta^{ab}$ and deconfined emergent boson and photon.
The shaded region denote the cross-over
to the conventional spiral phase which is confined.
}
\label{fig:c_b}
\end{figure}

\section{Possible spiral$^*$ phase}
In view of the finite size effect in the flux-flux correlation functions, 
it is desirable to do a systematic finite size scaling by increasing the size of lattice
and we increase the lattice size to $8^3 \times 16$ with $N=4$ fixed.
However, in the larger lattice it turns out that the Coulomb phase becomes 
unstable against a formation of spiral long range order with pitch $8$ 
in spatial direction.
In the smaller system with size $4$ in the spatial direction, 
the Coulomb phase was stabilized because
the spiral order with a pitch longer than the system size 
is suppressed by the periodic boundary condition. 
However, the presence of the long range order does not necessarily imply
the absence of fractionalization as we discussed earlier.
In the following, we examine the possibility of the coexistence between
the spiral order and fractionalization.

\begin{figure}
        \includegraphics[height=7cm,width=8cm]{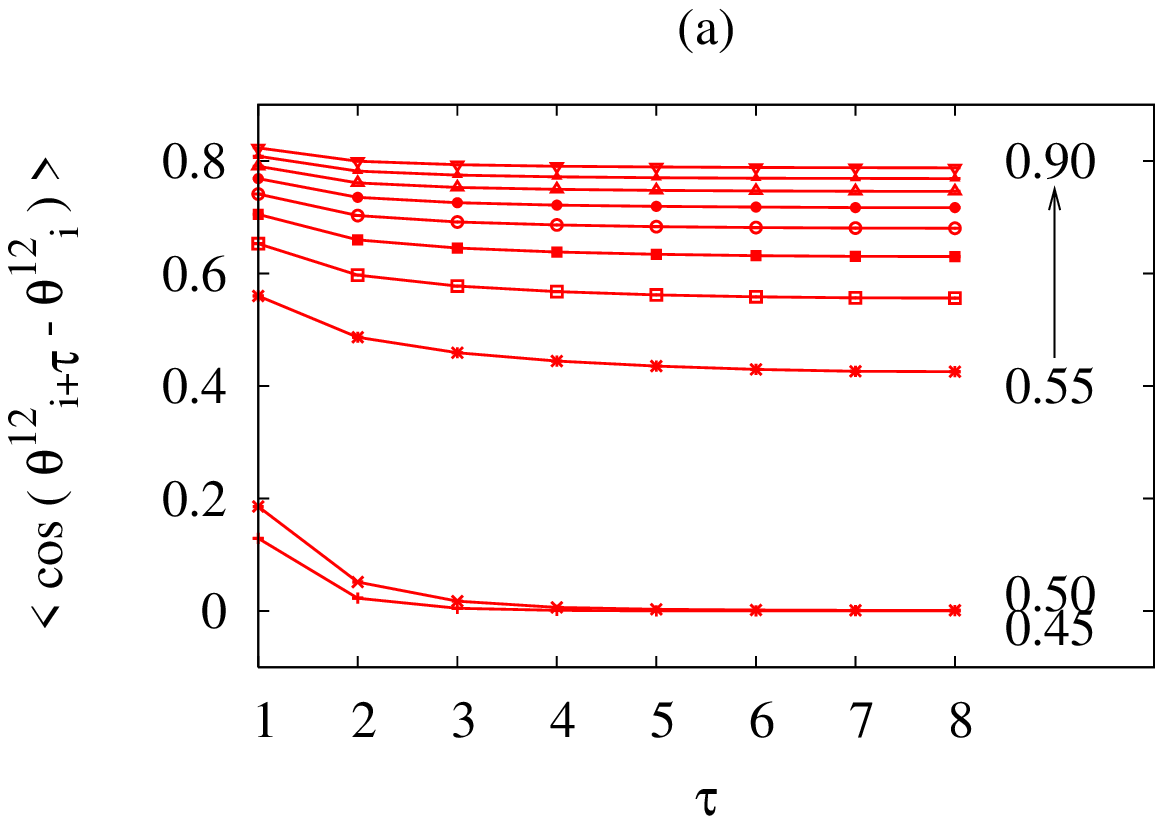}
        \includegraphics[height=7cm,width=8cm]{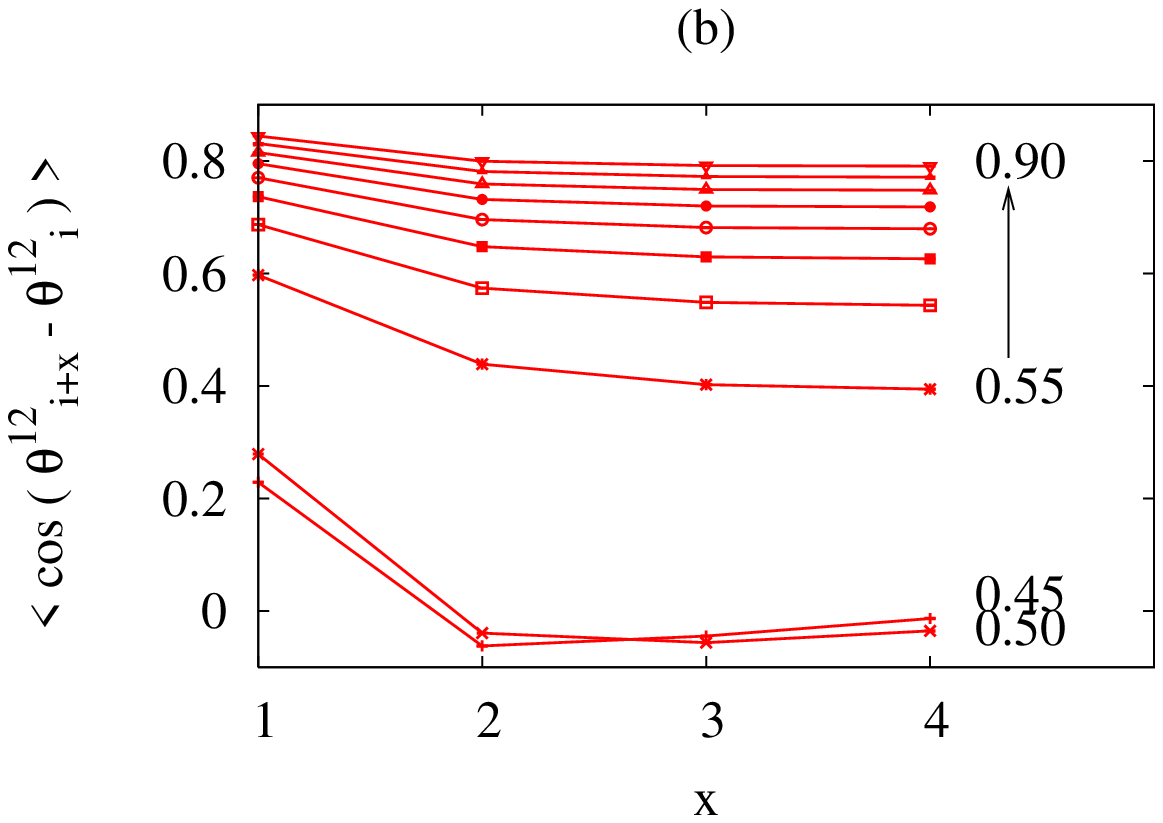}
        \includegraphics[height=7cm,width=8cm]{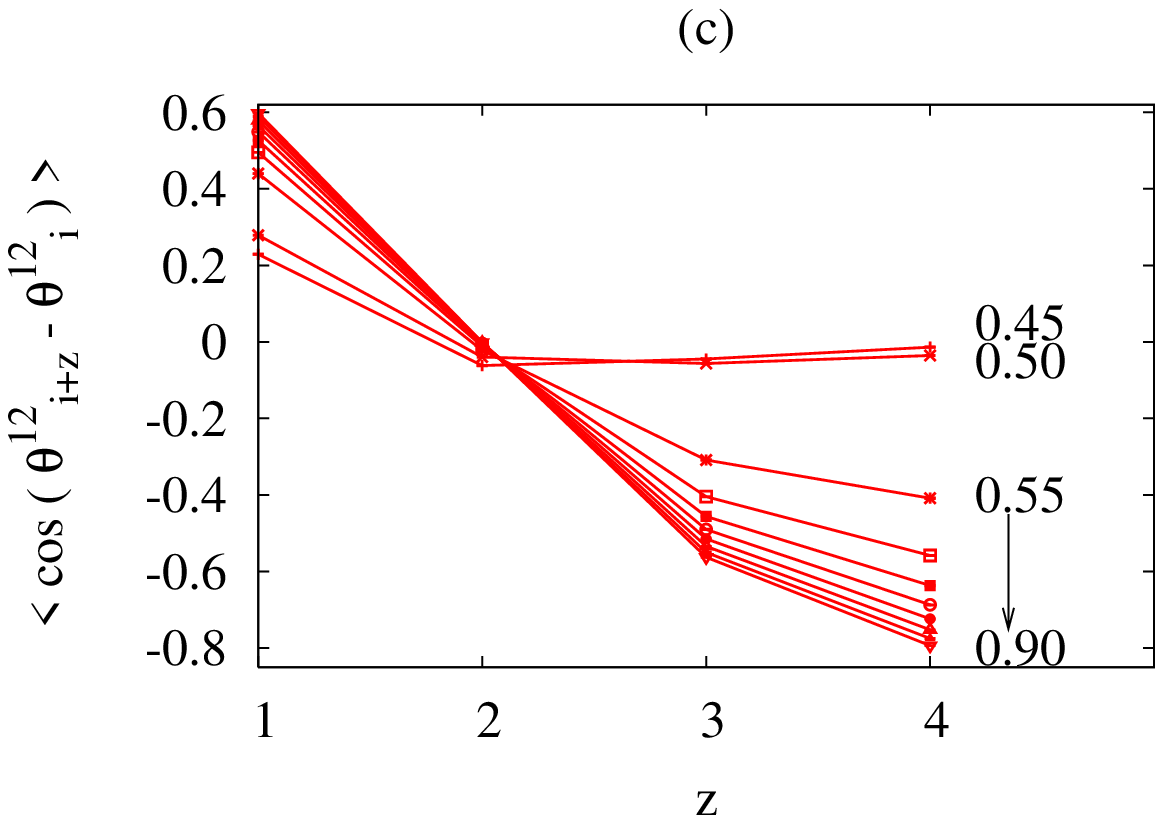}
\caption{
(color online) The correlation functions between $\theta^{12}$, the phases of exciton of flavor $1$ and $2$
in the temporal direction (a), the $x$-direction (b) and the $z$-direction (c).
The average is taken both for the ensemble and for the site $i$.
The numbers in the figure denote the value of $\kappa N$ for each curve. 
}
\label{fig:s_12}
\end{figure}

In Fig. \ref{fig:c_b}, we show the ``specific heat'' with frustration $\kappa^{'} N = -0.3$ 
in the lattice with size $8^3 \times 16$.
The time step is chosen such that $ \kappa_x/\kappa_0  = (1/0.6)^2$.
There is a peak in the ``specific heat'' around $\kappa N = 0.54$.
Although now shown here, there is a jump in the expectation value of the action between $\kappa N = 0.54$ and $0.55$,
signifying a first order phase transition.
The nature of the phases separated by the transition can be understood from 
the exciton correlation function which are displayed in Fig. \ref{fig:s_12}.
The temporal correlation and the spatial correlation in the $x$-direction
show the long range phase coherence for $\kappa N \geq 0.55$
as is shown in Fig. \ref{fig:s_12} (a) and (b).
On the other hand 
Fig. \ref{fig:s_12} (c) shows that
the $1,2$-exciton exhibits a spiral order in the $z$-direction 
for $\kappa N \geq 0.55$.
The pitch of the spiral order is $8$. 
However, it is likely that the pitch is not an intrinsic property but 
is the effect of the boundary condition of period $8$.
For an infinite system, it is possible to develop an incommensurate spiral order depending
on the coupling constants.
We proceed with the assumption that 
the nature of the spiral phase does not change greatly
even though the pitch may change in the thermodynamic limit.
Similar to the $1,2$-exciton, $ab$-exciton with different $a$, $b$ 
develops spiral order in certain directions.
This pattern of spiral order is summarized in Table \ref{tbl:1}.
In the second column of Table \ref{tbl:1} we show the off-diagonal excitons 
which exhibit spiral order in each direction.
It is convenient to understand the pattern of the spiral order 
in terms of the slave-boson field as is shown in the third column of Table \ref{tbl:1}.
For example, the spiral order of 
$\theta^{13}$, $\theta^{14}$, $\theta^{23}$ and $\theta^{24}$
in the $x$-direction
can be interpreted as the spiral order of $\phi^{1}$ and $\phi^{2}$ 
relative to $\phi^{3}$ and $\phi^{4}$.
It is emphasized that the individual slave-boson field can be 
either phase coherent or incoherent 
when the exciton fields have the long range phase coherence.
In other words, there can be two possible phases within the spiral phase.
One is the Bose condensed phase of the slave-bosons and 
the other, the pair Bose condensed phase 
without the Bose condensation of the individual slave-boson.
The former is the conventional spiral phase, 
while the latter is the fractionalized phase with emergent photon.
The latter possibility has been discussed 
in the uniform condensate called Higgs$^*$ phase\cite{LEE2}.
It can be understood 
as the condensation of bundles of 
vortices of individual slave-bosons
without the condensation of vortices of the individual slave-boson\cite{LEE2}.
Similar consideration applies to the spiral phase
and in order to distinguish the conventional spiral phase from the fractionalized spiral phase,
we will refer the latter phase as spiral$^*$ phase.

Unlike the Coulomb phase, it is very difficult to
unambiguously identify the spiral$^*$ phase, because
other gapless fluctuations are present in the form
of $N-1$ Goldstone modes.
It is necessary to minimize the contribution of the
Goldstone modes to the flux-flux correlation function.
For the reason we calculated the flux-flux correlation function 
defined as
\bq
C_W^{'}(r) = - Re \frac{1}{V} \sum_i \left< \delta W^{xy \dagger}_{1234}(i+r) \delta W^{xy}_{2143}(i) \right>,
\label{fx_p}
\eq
where $W^{\mu \nu}_{abcd}(i)$ is the Wilson loop operator defined in Eq. (\ref{W}).
Here we consider the flux-flux correlation function which is different from Eq. (\ref{fx})
and the correlation is calculated between two $W^{\mu \nu}_{abcd}$ 
whose flavor indices are ordered in a reverse way\cite{FLAVOR}.
This is to suppress the contribution of Goldstone modes 
in the presence of the spiral long range order. 
The Goldstone mode comes from the
$U(1)^{N-1}$ global symmetry under which
the phase of exciton transforms as 
$\theta^{ab}_i \rightarrow \theta^{ab}_i + \varphi^a - \varphi^b$.
Since the Wilson operator does not carry a net flavor, 
it is coupled to the Goldstone modes
only at a finite energy/momentum as
\bqa
\delta W^{\hat x \hat y}_{abcd} & \sim &
  \partial_x( \varphi^b - \varphi^d )
+ \partial_y( \varphi^c - \varphi^a ) \nn
& + & 
  \frac{1}{2} \partial_x^2( \varphi^b - \varphi^d )
+ \frac{1}{2} \partial_y^2( \varphi^c - \varphi^a )  + \cdots
\eqa
However, the first derivative term in the above expression does not contribute to $C^{'}_W(r)$
because of the different ordering of flavors in the two Wilson operators inside $C^{'}_W(r)$.
In energy-momentum space, the first derivative terms contribute 
to the temporal correlation function as 
$\int d^4k k_x k_y < \varphi^a_k \varphi^a_{-k} > e^{i k_0 \tau}$
$ \sim \int d^4k k_x k_y / k^2 e^{i k_0 \tau} $ and it vanishes. 
As a result, the first non-vanishing contribution from Goldstone modes come 
from the second derivative term and 
$C^{'}_W(k)$ is coupled to the Goldstone mode with an extra factor of $k^4$
(In contrast the correlation $C_W$ in Eq. (\ref{fx}) will only have a factor of $k^2$).
Upon Fourier transformation, the Goldstone mode contributes to $C^{'}_W(\tau)$ as $1/\tau^6$.
If we take into account finite size effect of spatial directions,
$C^{'}_W(\tau)$ can decay at most as $1/\tau^3$.
Thus the power-law decay of the temporal correlation function 
with a power smaller than $3$ can not be explained in terms of Goldstone mode.
On the other hand, the emergent photon mode will contribute to
$C^{'}_W(\tau)$ as $1/\tau^4$ in an infinite lattice 
just like the flux-flux correlation function in 3+1D electrodynamics.
The finite size effect can modify the behavior upto $1/\tau$.

\begin{figure}
        \includegraphics[height=7cm,width=8cm]{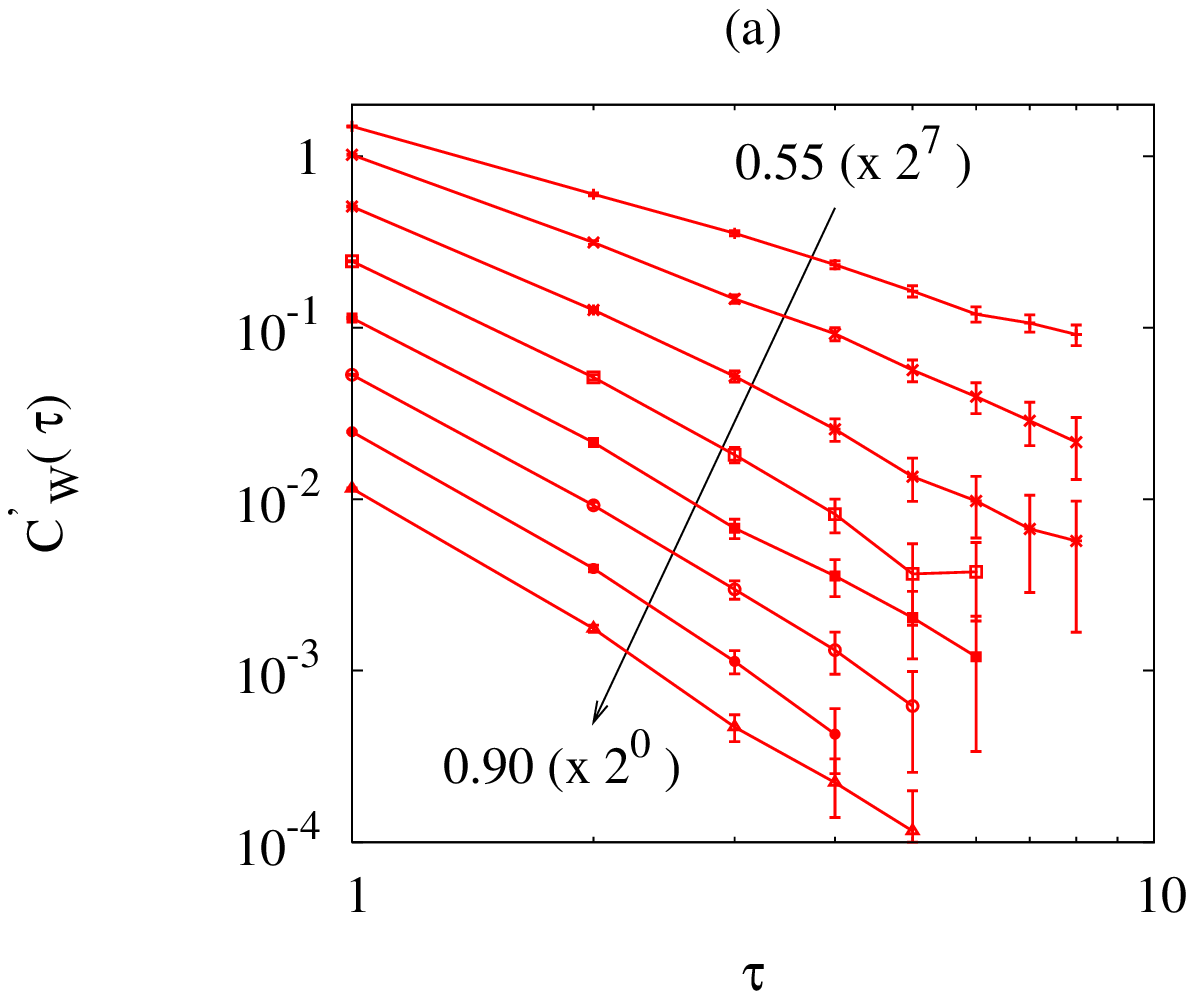}
        \includegraphics[height=7cm,width=8cm]{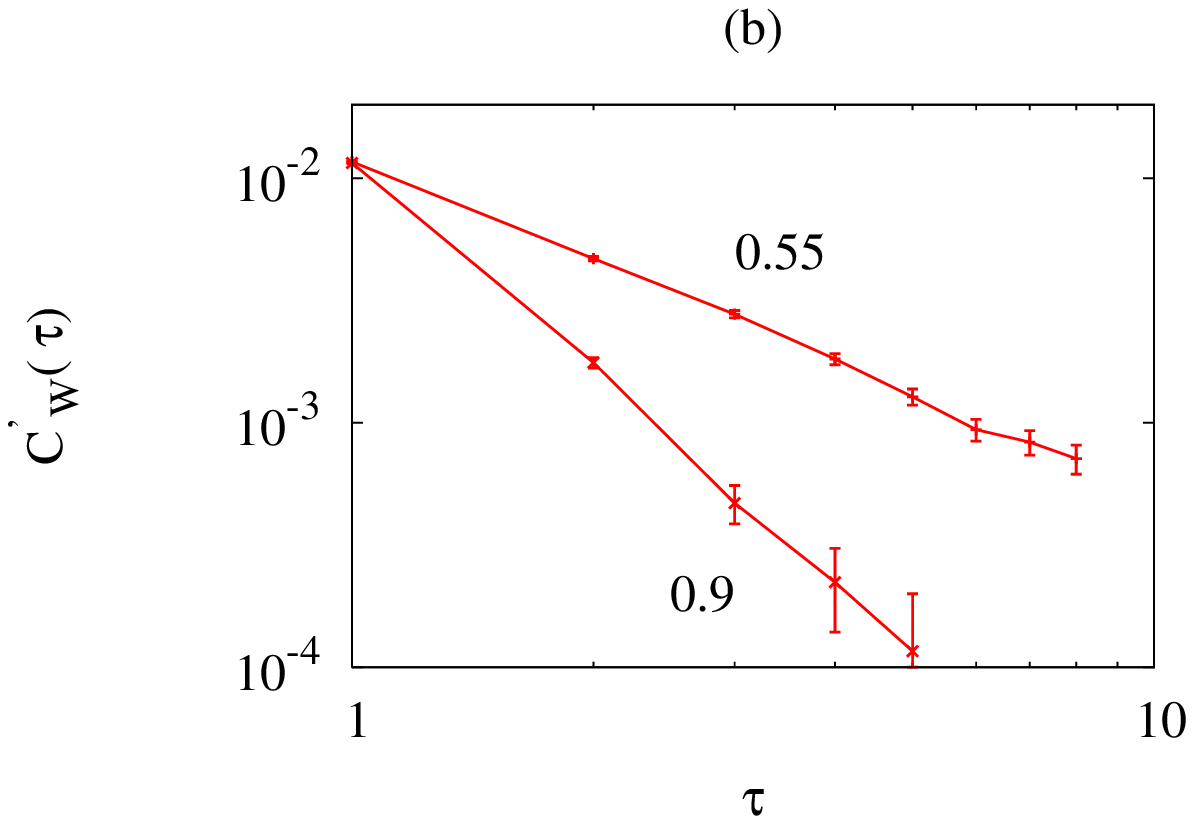}
        \includegraphics[height=7cm,width=8cm]{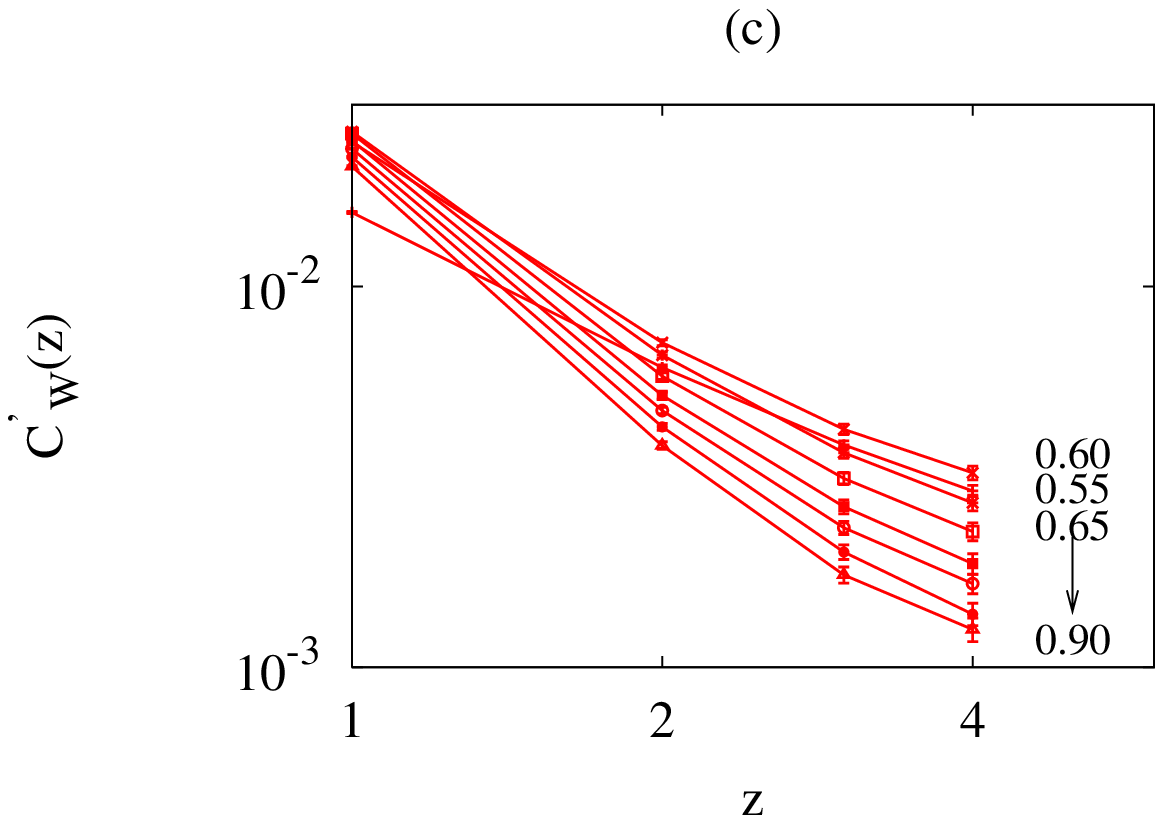}
\caption{
(color online) The correlations function $C^{'}_W$ between the fluctuations of Wilson operators.
(a) and (b) are for the temporal direction and (c) is for the $z$-direction.
In (a), different scale factors are multiplied to separate the curves as denoted in the figure.
In (b), the correlation functions are shown for $\kappa N = 0.55$ and $0.9$ with bare scale.
}
\label{fig:f}
\end{figure}

In the spiral phase the temporal flux-flux correlation function decays algebraically as is shown in Fig. \ref{fig:f} (a) and (b).
Result in the disordered phase is not shown because 
it is too small (less than $10^{-2}$ at nearest neighbor) and 
further neighbor correlation is completely dominated by statistical noises.
The decaying power of $C^{'}_W(\tau)$ is about $1.3$ for $\kappa N = 0.55$ 
and it increases with increasing $\kappa$.
The decaying power becomes about $3.3$ at $\kappa N = 0.9$ and $3.6$ at $\kappa N = 1$.
It first becomes greater than $3$ around $\kappa N = 0.85$.
Thus the spiral$^*$ phase may exist for $0.55 \leq \kappa N \leq 0.80$
and the conventional spiral phase, for $\kappa N \geq 0.85$, with a 
continuous phase transition in between.
However, it is a possibility 
that the slower decay for $\kappa N \leq 0.80$ 
is due to a cross-over behavior within the conventional spiral phase.
This is because the ``specific heat'' does not exhibit an anomaly 
within our error bar for the phase transition from the spiral to spiral$^*$ phases
as is shown in Fig. \ref{fig:c_b}.
In the spatial flux-flux correlation function,
the algebraic decay is less clear because of the finite size effect.
However, it is clear that the correlation decays more slowly for smaller $\kappa$ 
as is shown in Fig. \ref{fig:f} (c).
Although not shown here, the flux-flux correlation function has the same dipolar properties as 
$< b_z(r) b_z(0) >$, where $b_z$ is the z-component of the `magnetic' field in the emergent gauge theory.
As expected from the gauge theory, $C_W^{'}(r)$ is positive (negative) 
if $r$ is parallel to the z(x) -direction.

\section{Conclusion}
In conclusion, we observed the emergent photon 
in a meta-stable fractionalized phase
of the exciton Bose condensate.
In the Coulomb phase,
the photon mode arises as a gapless collective excitation 
although individual excitons are gapped.
The Coulomb phase is only meta-stable 
because it is unstable in a large lattice
and a spiral phase becomes stable.
In the spiral phase we studied a possibility 
where fractionalization 
coexists with the long range spiral order.
We observed a slow collective mode of exciton 
which can not be explained in terms of pure Goldstone modes
in the spiral phase and interpret the slow mode 
as the emergent photon in the fractionalized 
spiral (spiral$^*$) phase.

\section{Acknowledgement}
This work was supported by the NSF grant DMR-0517222. 
Part of the simulation was performed on computers at 
the National Energy Research Scientific Computing Center (NERSC).
We thank Peter Virnau, Olexei Motrunich, Matthias Troyer for their 
advices for Monte Carlo simulation, 
David Turner at NERSC for his help in parallel 
running of computer code, 
T. Senthil, Michael Hermele and Dmitri Ivanov for illuminating discussions, and
Ying Ran for his assistance with Linux system.
We also wish to thank the hospitality of the Aspen Center for Physics.


\newpage
\begin{thebibliography}{27}
\bibitem{LEE1} S.-S. Lee and P. A. Lee, Phys. Rev. B {\bf 72}, 235104 (2005).
\bibitem{BUTOV} L. V. Butov, A. C. Gossard and D. S. Chemla, Nature {\bf 418}, 751 (2002).
\bibitem{EXCEPTION} We are considering fractionalization in space dimension greater than one in the absence of magnetic field.
\bibitem{ANDERSON} P. W. Anderson, Science {\bf 235}, 1196 (1987); 
P. Fazekas and P. W. Anderson, Philos. Mag. {\bf 30}, 432 (1974).
\bibitem{KITAEV} A. Y. Kitaev, Ann. Phys. (N.Y.) {\bf 303}, 2 (2003).
\bibitem{WEN2003PRL} X.-G. Wen, Phys. Rev. Lett. {\bf 90}, 016803 (2003).
\bibitem{MOESSNER} R. Moessner and S. L. Sondhi, Phys. Rev. B {\bf 68}, 184512 (2003).
\bibitem{MOTRUNICH} O. I. Motrunich and T. Senthil, Phys. Rev. Lett. {\bf 89}, 277004 (2002).
\bibitem{WEN2002PRL} X.-G. Wen, Phys. Rev. Lett. {\bf 88}, 11602 (2002).
\bibitem{MOTRUNICH2004} O. I. Motrunich and T. Senthil, Phys. Rev. B {\bf 71}, 125102 (2005).
\bibitem{HERMELE} M. Hermele, M. P. A. Fisher and L. Balents, Phys. Rev. B {\bf 69}, 064404 (2004).
\bibitem{LEVIN} M. A. Levin and X.-G. Wen, Phys. Rev. B {\bf 67}, 245316 (2003).
\bibitem{SENTHIL1} T. Senthil and M. P. A. Fisher, Phys. Rev. B {\bf 63}, 134521 (2001); T. Senthil, M. Vojta, and S. Sachdev, Phys. Rev. B {\bf 69}, 035111 (2004).
\bibitem{LEE2} S.-S. Lee, T. Senthil and P. A. Lee, submitted to Phys. Rev. B (cond-mat/0509380).
\bibitem{CRUETZ} M. Creutz, L. Jacobs and C. Rebbi, Phys. Rev. D {\bf 20}, 1915 (1979). 
\bibitem{FLAVOR} 
In Ref. \cite{LEE1}, all flavor indices are summed 
in the definition of the Wilson operator. 
Here, we will consider the case where the flavor symmetry is broken 
and the different flavor combination should be weighted
in different ways.
Since any flavor combination will couple to the gauge field,
we consider one specific combination, i.e., $1,2,3,4$.
Owing to the spiral order, the correlation function between
$\delta W^{xy \dagger}_{1234}(i+r)$ and $\delta W^{xy}_{2143}(i)$ 
is negative in the temporal direction.
This implies that $W^{xy }_{1234}$ and $\delta W^{xy}_{2143}$ contribute
to the flux operator of the gauge theory with opposite signs.
Thus we added an additional minus sign in the definition of flux-flux correlation function in Eq. (\ref{fx_p}).




\end{thebibliography}
\end{document}